\documentclass[aps,preprint,showpacs,floats,epsf,epsfig,nofootinbib,12pt]{revtex4}
\usepackage{graphicx}% Include figure files
\usepackage{epsfig}
\usepackage[dvips]{color}

\def\beq{\begin{eqnarray}}
\def\eeq{\end{eqnarray}}
\def\non{\nonumber}

\def\la{\langle}
\def\ra{\rangle}

\begin{document}

\title{Study on the mixing of $\Xi_c$ and  $\Xi'_c$  by the transition $\Xi_{b}\to\Xi^{(')}_c$}

\vspace{1cm}

\author{ Hong-Wei Ke$^{1}$\footnote{khw020056@tju.edu.cn}, Gang-Yang Fang$^{1}$ and
     Yan-Liang Shi$^2$\footnote{ys6085@princeton.edu}
   }

\affiliation{  $^{1}$ School of Science, Tianjin University, Tianjin 300072, China
\\
  $^{2}$ Princeton Neuroscience Institute, Princeton University, Princeton, NJ 08544, USA}

\vspace{12cm}

\begin{abstract}
Recently, the LHCb collaboration has observed the decays
$\Xi^0_{b}\to\Xi^+_{c}D^-_s$ and $\Xi^-_{b}\to\Xi^0_{c}D^-_s$.
They measured the relative branching fractions times the ratio of
beauty-baryon production cross-sections
$\mathcal{R}(\frac{\Xi^0_b}{\Lambda_b})\equiv
\frac{\sigma(\Xi_b^0)}{\sigma(\Lambda^0_b)}\times\frac{B(\Xi^0_{b}\to\Xi^+_{c}D^-_s)}{B(\Lambda^0_{b}\to\Lambda^+_{c}D^-_s)}$
and $\mathcal{R}(\frac{\Xi^-_b}{\Lambda_b})\equiv
\frac{\sigma(\Xi^-_b)}{\sigma(\Lambda^0_b)}\times\frac{B(\Xi^-_{b}\to\Xi^0_{c}D^-_s)}{B(\Lambda^0_{b}\to\Lambda^+_{c}D^-_s)}$.
Once the ratio $\frac{\sigma(\Xi_b^0)}{\sigma(\Lambda^0_b)}$ or $\frac{\sigma(\Xi_b^-)}{\sigma(\Lambda^0_b)}$  is known, one can determine the  relative branching fractions which can be used to exam the mixing of $\Xi_c$ and  $\Xi'_c$. In previous literature, $\Xi_c$ and  $\Xi'_c$ were assumed to belong to SU(3)$_F$ antitriple and sextet, respectively. However,  recent experimental measurements such as  the ratio $\Gamma(\Xi_{cc}\to\Xi_c\pi^+)/\Gamma(\Xi_{cc}\to\Xi'_c\pi^+)$ indicate the spin-flavor structures of $\Xi_{c}$ and $\Xi'_{c}$ are a mixture  of $\Xi^{\bar 3}_{c}$ and $\Xi^{6}_{c}$. The exact value of mixing angle $\theta$ is still under debate.  In theoretical models, the mixing angle was fitted to be about $16.27^\circ\pm2.30^\circ$ or $85.54^\circ\pm2.30^\circ$ based on decay channels  $\Xi_{cc} \to\Xi^{(')}_{cc} $.  While in lattice calculation,  a small angle ($1.2^\circ\pm0.1^\circ$) is preferred.  To address such discrepancy and test the mixing of $\Xi_c$ and  $\Xi'_c$, here we propose the analysis of semileptonic and non-leptonic decays of  $\Xi_{b}\to\Xi_{c}$ and $\Xi_{b}\to\Xi^{'}_{c}$. We calculate the decay rate of  $\Xi_{b}\to\Xi_{c}$ and $\Xi_{b}\to\Xi^{'}_{c}$ based on light-front quark model and study the effect of the mixing angle on the ratios of weak decays $\Xi_{b}\to\Xi_{c}$ and $\Xi_{b}\to\Xi^{'}_{c}$.  In particular, we find the transition $\Xi_{b}\to\Xi^{'}_{c}$ can be an ideal channel to verify the mixing and extract the mixing angle because in theory the decay rate would be extremely tiny without mixing. Our calculation suggests a measurement of $\Xi_{b}\to\Xi'_{c}$ can be feasible in the near future, which will help to test flavor mixing angle  $\theta$ and elucidate the mechanism of decay of heavy baryons.

\end{abstract}

\maketitle

\section{Introduction}

Recently the LHCb collaboration observed the decays
$\Xi^0_{b}\to\Xi^+_{c}D^-_s$ and $\Xi^-_{b}\to\Xi^0_{c}D^-_s$\cite{LHCb:2023ngz}.
They measured the relative branching fractions times the ratio of the corresponding
beauty-baryon production cross-sections
$\mathcal{R}(\frac{\Xi^0_b}{\Lambda_b})\equiv
\frac{\sigma(\Xi_b^0)}{\sigma(\Lambda^0_b)}\times\frac{B(\Xi^0_{b}\to\Xi^+_{c}D^-_s)}{B(\Lambda^0_{b}\to\Lambda^+_{c}D^-_s)}=15.8\pm1.1\pm0.6\pm7.7$
and $\mathcal{R}(\frac{\Xi^-_b}{\Lambda_b})\equiv
\frac{\sigma(\Xi^-_b)}{\sigma(\Lambda^0_b)}\times\frac{B(\Xi^-_{b}\to\Xi^0_{c}D^-_s)}{B(\Lambda^0_{b}\to\Lambda^+_{c}D^-_s)}=16.9\pm1.3\pm0.9\pm4.3$, which suggests experimental measurement  of transitions $\Xi_{b}\to\Xi_{c}$ and $\Xi_{b}\to\Xi'_{c}$ can be promising in the near future. These channels can be used to probe the possible mixing of $\Xi_{c}$  and $\Xi'_{c}$ states.

Historically,  $\Xi_c$ and  $\Xi'_c$ were identified as SU(3)$_F$ antitriple and sextet of flavor symmetry, respectively. For these kinds of heavy baryons, usually the two light quarks $q$ (denotes $u$ or $d$) and $s$ are assumed to form a subsystem called
diquark \cite{Ebert:2006rp,Korner:1992wi}.  In previous literature, a presupposition is  the spin-flavor structure of $\Xi_{c}$ is $[qs]_0 c$
 and that of  $\Xi^{'}_{c}$ is $[qs]_1 c$ where the subscript 0 or 1 represents the total spin of the $qs$ subsystem.

However, recent experimental data such as  the ratio of the branching fraction
$\Xi_{cc}\to\Xi'_c\pi^+$ relative to that of the decay
$\Xi_{cc}\to\Xi_c\pi^+$ \cite{LHCb:2022rpd}  indicate the spin-flavor structures of $\Xi_{c}$ and $\Xi'_{c}$ are not pure  $[qs]_0 c$ and $[qs]_1 c$  (they are denoted as $\Xi^{\bar 3}_{c}$ and $\Xi^{6}_{c}$, respectively, where the superscripts $\bar 3$ and $6$  correspond to SU(3)$_F$ anti-triple and sextet) but mixtures of them. In this case,  we can use a mixing angle $\theta$ ($0< \theta<\pi$) to describe the composition of  $\Xi^{\bar 3}_{c}$ and $\Xi^{6}_{c}$ in flavor-spin wave functions of $\Xi_{c}$  and $\Xi'_{c}$, i.e., $| \Xi_{c} \ra =  {\rm cos}\theta\, | \Xi^{\bar 3}_{c} \ra + {\rm sin}\theta\, | \Xi^{6}_{c} \ra$ and $| \Xi'_{c} \ra= -{\rm sin}\theta\, | \Xi^{\bar 3}_{c} \ra+{\rm cos}\theta\, | \Xi^{6}_{c} \ra$. When $\theta$ is set to 0, i.e.,  $\Xi_{c}=\Xi^{\bar 3}_{c}$ and $\Xi^{'}_{c}= \Xi^{6}_{c}$,  the structures of $\Xi_{c}$ and $\Xi^{'}_{c}$ restore the original setting supposed by the authors of Ref. \cite{,Korner:1992wi,Ebert:2006rp}.  Such mixing is a reflection of the breaking of SU(3) flavor symmetry, which is due to the mass difference between $s$ and $u,d$ quarks.

To determine the mixing angle between  $\Xi_{c}$ and $\Xi^{'}_{c}$, different approaches has been proposed while a consensus has not been achieved. One approach is to analyze transition rates of these state based on phenomenological models.  In Ref. \cite{Ke:2022gxm},  we fixed the mixing angle to be $16.27^\circ\pm2.30^\circ$ or $85.54^\circ\pm2.30^\circ$ by  fitting the data $\Gamma(\Xi_{cc}\to\Xi_c^{'+}\pi^+)/\Gamma(\Xi_{cc}\to\Xi_c\pi^+)$. Geng et. al. \cite{Geng:2022yxb,Geng:2022xfz} obtain the value is $24.66^\circ$ from the mass spectra. Another approach is lattice simulation.  Recent lattice calculation \cite{Liu:2023feb} prefers   $1.2^\circ\pm0.1^\circ$, which is much smaller than those from phenomenological analysis.

To better understand the mixing mechanism and obtain a precise measurement of mixing angle of $\Xi_{c}$ and $\Xi^{'}_{c}$,   more experimental data involving $\Xi_{c}$ and $\Xi'_{c}$ are needed. Inspired by recent data from LHCb on $\Xi_{b}\to\Xi_{c}D_s$, here we apply light-front quark model \cite{Jaus,Ji:1992yf,Cheng:1996if,Cheng:2003sm,Lu:2007sg,
	Li:2010bb,Ke:2009ed,Ke:2010,Choi:2007se,
	Ke:2009mn,Ke:2011fj,Ke:2011mu,Ke:2011jf,Chua:2018lfa,Yu:2017zst,
	Ke:2017eqo,pentaquark1,pentaquark2,Tawfiq:1998nk,Chang:2018zjq,Geng:2022xpn} framework to study the  semileptonic and non-leptonic decays of  $\Xi_{b}\to\Xi_{c}$ and $\Xi_{b}\to\Xi'_{c}$, and estimate the feasibility to extract  the mixing angle  from these channels.

The light-front quark model has been widely used to study light-quark systems. In our earlier work we extended the light-quark model to study baryon with heavy quark-diquark picture \cite{Ke:2007tg,Ke:2012wa,Wei:2009np}. Later the three-quark picture of baryon was adopted \cite{Ke:2019smy,Ke:2019lcf,Lu:2023rmq}. In this study we also apply the three-quark picture to analyze decays of heavy-baryon. In the calculation, we apply the spectator approximation to the light diquark components.

This paper is
organized as follows: after the introduction, in section II we discuss the form factors for the transition $\Xi_{b}\rightarrow \Xi^{(')}_{c}$
in the light-front quark model. Our numerical results for $\Xi_{b}\rightarrow
\Xi^{(')}_{c}$ are presented in section
III. The section IV is devoted to our conclusion and discussions.

\section{the  form factors of $\Xi_{b}\to\Xi_c$ and $\Xi_{b}\to\Xi^{'}_c$ in LFQM}

\begin{figure}
\begin{center}
\begin{tabular}{ccc}
\scalebox{0.6}{\includegraphics{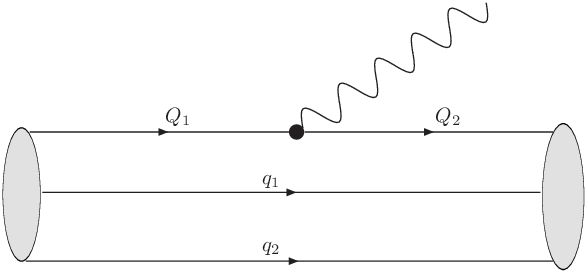}}
\end{tabular}
\end{center}
\caption{The Feynman diagrams for $\Xi_{b}\to\Xi^{(')}_{c}$ transitions,
where $Q_1$ and $Q_2$ denote heavy quark in the initial and final states, respectively. $q_1$ and $q_2$ represent light quark states. $\bullet$ denotes $V-A$ current vertex.}\label{t1}
\end{figure}

The leading Feynman diagram responsible for the weak decay
$\Xi_{b}\to\Xi_c^{(')}$ is shown in Fig. \ref{t1}.
During this process,  $b$ quark decays to $c$ quark and diquark $sq$ stands as the spectator. Here spectator approximation is directly applied to diquark state $sq$ because its spin configuration does not change during the transition at the leading order, which greatly alleviates the theoretical difficulties for calculating the hadronic transition matrix elements.

The form factors for the weak transition $\Xi_{b}\rightarrow
\Xi_{c}$  are defined  as
\begin{eqnarray}\label{s1}
&& \la \Xi_{c}(P',S',S_z') \mid \bar{Q'}\gamma_{\mu}
 (1-\gamma_{5})Q\mid \Xi_{b}(P,S,S_z) \ra  \non \\
 &=& \bar{u}_{\Xi_{c}}(P',S'_z) \left[ \gamma_{\mu} f_{1}(q^{2})
 +i\sigma_{\mu \nu} \frac{ q^{\nu}}{M_{\Xi_{b}}}f_{2}(q^{2})
 +\frac{q_{\mu}}{M_{\Xi_{b}}} f_{3}(q^{2})
 \right] u_{\Xi_{b}}(P,S_z) \nonumber \\
 &&-\bar u_{\Xi_{c}}(P',S'_z)\left[\gamma_{\mu} g_{1}(q^{2})
  +i\sigma_{\mu \nu} \frac{ q^{\nu}}{M_{\Xi_{b}}}g_{2}(q^{2})+
  \frac{q_{\mu}}{M_{\Xi_{b}}}g_{3}(q^{2})
 \right]\gamma_{5} u_{\Xi_{b}}(P,S_z).
\end{eqnarray}
where  $P$ and $P'$ denote four-momentum of  $ \Xi_{b}$ and $\Xi_{c}$, respectively. $q \equiv P-P'$.

In terms of the spin-flavor structures of $\Xi_{b}$ and $\Xi_{c}$, we assumed  $| \Xi_{c} \ra $ is a mixture of state $ | \Xi^{	\bar 3}_{c} \ra $ and $| \Xi^{6}_{c} \ra$, with a mixing angle $\theta$ \cite{Ke:2022gxm}:
  \begin{equation}
 | \Xi_{c} \ra   =   {\rm cos}\theta\,  | \Xi^{	\bar 3}_{c} \ra +{\rm sin} \theta\, | \Xi^{6}_{c} \ra   \ ,
 \end{equation}
 and for $ | \Xi^{'}_{c} \ra$:
 \begin{equation}
 | \Xi^{'}_{c} \ra= -{\rm sin}\theta\, | \Xi^{	\bar 3}_{c} \ra +{\rm cos}\theta\,  | \Xi^{6}_{c}  \ra  \  .
 \end{equation}

 With heavy-quark symmetry, we can also express $| \Xi_{b} \ra $ as a superposition of flavor eigenstates:

\begin{equation}
 | \Xi_{b} \ra   =   {\rm cos}\theta\,  | \Xi^{ \bar	3}_{b} \ra +{\rm sin} \theta\, | \Xi^{6}_{b} \ra \ ,
 \end{equation}
 Here we take the limit of heavy-quark symmetry, hence the mixing angle $\theta$ is the same for $b$-quark baryons.

Using  eigenstates of $\text{SU(3)}_F$,  $ | \Xi^{6}_{b} \ra $,  $ | \Xi^{ \bar	3}_{b} \ra $, $| \Xi^{	\bar 3}_{c} \ra$, $| \Xi^{6}_{c} \ra  $,  the matrix element $\la \Xi_{c}(P',S',S_z') \mid \bar{c}\gamma_{\mu}
 (1-\gamma_{5})b \mid \Xi_{b}(P,S,S_z) \ra$ can be written as
\begin{equation}\label{mixing}
{\rm cos}^2\theta\la \Xi^{\bar 3}_{c}
\mid \bar{c}\gamma_{\mu}
 (1-\gamma_{5})b \mid \Xi^{\bar 3}_{b} \ra+{\rm sin}^2\theta\la \Xi^{6}_{c} \mid
\bar{c}\gamma_{\mu}
 (1-\gamma_{5})b \mid \Xi^{ 6}_{b} \ra.
\end{equation}
 where we neglect the transition matrix elements for $\Xi^{\bar 3}_{b}\to\Xi^{6}_c$ and $\Xi^{6}_{b}\to\Xi^{\bar 3}_c$ because they are forbidden in the leading-order approximation of light-front quark model.  In certain phenomenological approaches the transitions $\Xi^{\bar 3}_{b}\to\Xi^{6}_c$ and $\Xi^{6}_{b}\to\Xi^{\bar 3}_c$ maybe exist in higher order of $\alpha_s$  but they should be  very suppressed like the decays of $\Lambda_{b}\to\Sigma_c$.

   For the transition matrix elements $\la \Xi^{\bar 3}_{c} \mid \bar{s}\gamma_{\mu}  (1-\gamma_{5})c \mid \Xi^{\bar 3}_{b}\ra$, the corresponding form factors are denoted as $f_i^s$, $g_i^s$ (See Eq. (\ref{s1}) for the definition of form factors).  Similarly, for $\la \Xi^{6}_{c}
    \mid \bar{s}\gamma_{\mu}   (1-\gamma_{5})c \mid \Xi^{6}_{b}\ra$, the form factors are denoted as $f_i^v$ , $g_i^v$. Combining Eq. (\ref{s1}) and Eq. (\ref{mixing}),  we have
\begin{eqnarray}\label{relation}
f_i={\rm cos}^2\theta f^s_i+{\rm sin}^2\theta f^v_i,\nonumber\\
g_i={\rm cos}^2\theta g^s_i+{\rm sin}^2\theta g^v_i.
\end{eqnarray}

 Following the
procedures given in
Refs. \cite{pentaquark1,pentaquark2,Ke:2007tg,Ke:2012wa}, the
transition matrix element can be computed using light-front approach with specific forms of vertex functions of $\Xi^{\bar 3}_{b}$, $\Xi^{\bar 3}_{c}$, $\Xi^{6}_{b}$ and $\Xi^{6}_{c}$. The calculations for $\Xi^{\bar 3}_{b}\to\Xi^{\bar 3}_{c}$ and $\Xi^{6}_{b}\to\Xi^{6}_{c}$ are same as those for $\Lambda_b\to\Lambda_c$ and $\Sigma_b\to\Sigma_c$, respectively.
The detailed derivation and the expressions $f_i^s$, $g_i^s$, $f_i^v$,
$g_i^v$ can be found in our earlier paper \cite{Ke:2019smy} .

For the transition matrix element $\la \Xi^{'}_{c}(P',S',S_z') \mid
\bar{c}\gamma_{\mu}
 (1-\gamma_{5})b \mid \Xi_{b}(P,S,S_z) \ra$,  by using   $ | \Xi^{6}_{b} \ra $,  $ | \Xi^{ \bar	3}_{b} \ra $, $| \Xi^{	\bar 3}_{c} \ra$, $| \Xi^{6}_{c} \ra  $,  it can be written as
 \begin{equation}\label{mixingp}
 -{\rm cos} \theta {\rm sin} \theta\la \Xi^{\bar 3}_{c}
 \mid \bar{c}\gamma_{\mu}
 (1-\gamma_{5})b \mid \Xi^{\bar 3}_{b} \ra+  {\rm cos} \theta {\rm sin} \theta
 \la \Xi^{6}_{c} \mid
 \bar{c}\gamma_{\mu}
 (1-\gamma_{5})b \mid \Xi^{ 6}_{b} \ra.
 \end{equation}
 The form factors of transition matrix element are also defined as in Eq. (\ref{s1}). Here we
just add a symbol `` $'$ " on $f_1$, $f_2$, $g_1$ and $g_2$  to
distinguish the quantities for $\Xi_{b}\to\Xi'_c$ and those for
$\Xi_{b}\to \Xi_c$. They are given by
\begin{eqnarray}\label{relation1}
f'_i={\rm cos}\theta {\rm sin}\,\theta(f^v_i- f^s_i),\nonumber\\
g'_i={\rm cos}\theta {\rm sin}\,\theta( g^v_i -g^s_i).
\end{eqnarray}

\section{Numerical Results}

\subsection{The results for $\Lambda_{b}\to \Lambda_c$}

In Ref. \cite{Ke:2019smy} we extended three-quark picture to study the decay of $\Lambda_b\to \Lambda_c$ and $\Sigma_b\to \Sigma_c$.
In that paper we employ a pole form in Eq. (\ref{p1}) to fix the form factor of the transition.

 \begin{eqnarray}\label{p1}
 F(q^2)=\frac{F(0)}{\left(1-\frac{q^2}{M_{\mathcal{B}_{i}}^2}\right)
  \left[1-a\left(\frac{q^2}{M_{\mathcal{B}_{i}}^2}\right)
  +b\left(\frac{q^2}{M_{\mathcal{B}_{i}}^2}\right)^2\right]},
 \end{eqnarray}
where $M_{\mathcal{B}_{i}}$ is the mass of the initial baryon.

%\begin{table}
%\caption{Quark mass  (in units of
% GeV).}\label{Tab:t1}
%\begin{ruledtabular}
%\begin{tabular}{cccc}
% $m_b$& $m_c$  & $m_s$  &$m_{u}$ \\\hline
% 4.64& $1.3$  & $0.37$  & 0.26
%\end{tabular}
%\end{ruledtabular}
%\end{table}

\begin{figure}[hhh]
\begin{center}
\scalebox{0.7}{\includegraphics{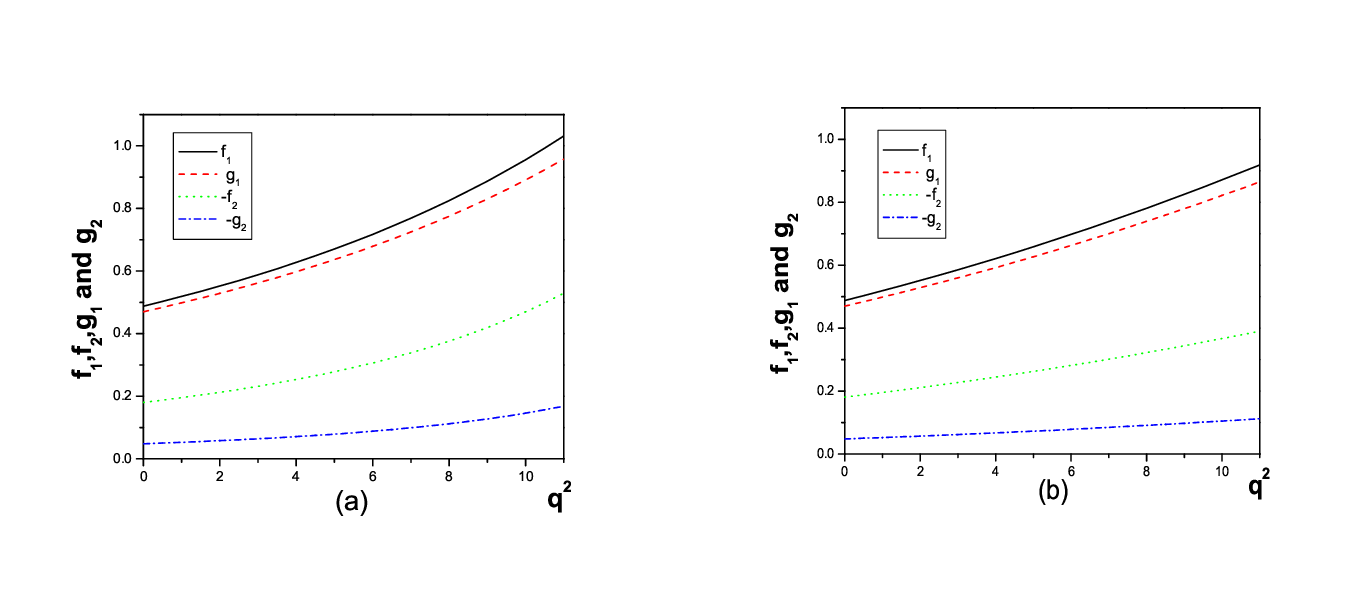}}
\end{center}
\caption{Form factors for the decay
$\Lambda_{c} \to \Lambda_{c}$ (a) pole form and (b) polynomial form.}\label{fgl}
\end{figure}

Later we find it doesn't work for the decay of double charmed baryon so in Refs. \cite{Ke:2019lcf,Lu:2023rmq} we employed a polynomial to parameterize these form factors  $f^s_i$, $g^s_i$, $f^v_i$ and $g^v_i$ ($i=1,2$),
 \begin{eqnarray}\label{p2}
 F(q^2)=F(0)\left[1+a\left(\frac{q^2}{M_{\mathcal{B}_{i}}^2}\right)
  +b\left(\frac{q^2}{M_{\mathcal{B}_{i}}^2}\right)^2\right].
 \end{eqnarray}
In this paper we first check whether the above polynomial formula also can be used for the transition $\Lambda_b\to\Lambda_c$. With the same parameters (quark masses and $\beta$ parameters in wave functions) in Ref. \cite{Ke:2019smy} we refit the parameters in Eq. (\ref{p2}).
The fitted values of $a,~b$ and $F(0)$ in the form factors
are presented in Table \ref{fgls}. We also depict two results in Fig. \ref{fgl}. The left graph (a) is the result from Ref.\cite{Ke:2019smy} and the right one (b) is the result this time.
Comparing these plots, one can find they are nearly equal to each other.
With these form factors, we re-calculate the decay rates  $\Gamma$, the integrated longitudinal and transverse asymmetries $a_l$ and $a_t$ and their ratio $R=a_l/a_t$ of
semi-leptonic for the transtion $\Lambda_b\to \Lambda_c$ and the results are present in table. \ref{lblcl}.
We also collect the results in Refs. \cite{Ke:2007tg,Ke:2019smy} for comparison. From the values in Tab. \ref{lblcl}, we can conclude that two extrapolative forms of the form factors don't significantly affect the results of transitions of $\Lambda_b\to \Lambda_c$. We also reevaluate the width and up-down asymmetry $\alpha$ of several no-leptonic decay channels. The results of the nonleptonic decays in table \ref{lblcn} are also very consistent.  For the detailed  expressions used in the calculation of these physical quantities,  one can refer to Refs. \cite{Cheng:1996cs,Korner:1994nh,Korner:1991ph,Bialas:1992ny}
and appendix of the Ref. \cite{Ke:2019smy}.

\begin{table}
\caption{The form factors given in
 polynomial form for the transition $\Lambda_b\to \Lambda_c$.}\label{fgls}
\begin{ruledtabular}
\begin{tabular}{ccccc}
  $F$    &  $F(0)$ &  $a$  &  $b$\\\hline
  $f^s_1$  &   0.488    & 1.94    &1.70   \\
$f^s_2$  &   -0.181     &2.44  &2.52   \\
  $g^s_1$  &      0.470   &    1.86& 1.58  \\
  $g^s_2$  &      -0.048  &    2.75  &  3.07
\end{tabular}
\end{ruledtabular}
\end{table}

\begin{table}
\caption{The widths (in unit of $10^{10}{\rm s}^{-1}$) and polarization asymmetries of $\Lambda_b\to
\Lambda_c l\bar{\nu}_l$ .}\label{lblcl}
\begin{ruledtabular}
\begin{tabular}{c|ccccc}
    &  $\Gamma$ &  $a_L$  &  $a_T$   & $R$    &  $P_L$ \\\hline
  The results in \cite{Ke:2019smy} & 4.22&-0.962&-0.766&1.54&-0.885  \\\hline
  The results in \cite{Ke:2007tg}
    &  5.15        &   -0.932 & -0.601 & 1.47 & -0.798 \\\hline
 This work  & 4.36       &   -0.933 & -0.668 & 1.49 & -0.826
  \end{tabular}
\end{ruledtabular}
\end{table}

\begin{table}
\caption{The widths (in unit of $10^{10}{\rm s}^{-1}$) and up-down asymmetries of non-leptonic decays
$\Lambda_b\to\Lambda_c M$.}\label{lblcn}
\begin{ruledtabular}
\begin{tabular}{c|cc|cc|cc}
  & \multicolumn{2}{c|}{the results in Ref. \cite{Ke:2019smy}~~~~~~}
  & \multicolumn{2}{c|}{ the results in \cite{Ke:2007tg}~~~~} &\multicolumn{2}{c}{ this work ~~~~}
\\\hline
  & $\Gamma$ & $\alpha$ & $\Gamma$  & $\alpha$  & $\Gamma$  & $\alpha$          \\\hline
  $\Lambda_b^0\to\Lambda_c^+ \pi^-$  & $0.261$  & $-0.999$
                        & $0.307$  & $-1$  & $0.261$  & $-0.999$    \\\hline
  $\Lambda_b^0\to\Lambda_c^+ \rho^-$ & $0.769$  & $-0.875$
                        & $0.848$  & $-0.883$   & $0.774$  & $-0.875$\\\hline
  $\Lambda_b^0\to\Lambda_c^+ K^-$    & $0.0209$  & $-0.999$
                        & $0.0247$  & $-1$ & $0.0210$  & $-0.999$     \\\hline
  $\Lambda_b^0\to\Lambda_c^+ K^{*-}$  & $0.0398$  & $-0.836$
                        & $0.0440$  & $-0.846$& $0.0402$  & $-0.837$  \\\hline
  $\Lambda_b^0\to\Lambda_c^+ a_1^-$  & $0.758$  & $-0.710$
                        & $0.838$  & $-0.726$ & $0.770$  & $-0.710$ \\\hline
  $\Lambda_b^0\to\Lambda_c^+ D_s^-$  & $0.927$ & $-0.974$
                        & $0.932$ & $-0.982$  & $0.928$ & $-0.987$ \\\hline
  $\Lambda_b^0\to\Lambda_c^+ D^{*-}_s$& $1.403$ & $-0.327$
                        & $1.566$  & $-0.360$ & $1.446$ & $-0.330$ \\\hline
  $\Lambda_b^0\to\Lambda_c^+ D^-$  & $0.0355$ & $-0.979$
                        & $0.0410$  & $-0.986$ & $0.0357$ & $-0.989$ \\\hline
  $\Lambda_b^0\to\Lambda_c^+ {D^*}^-$& $0.0630$& $-0.371$
                        & $0.0702$  & $-0.403$& $0.0649$& $-0.374$
\end{tabular}
\end{ruledtabular}
\end{table}

\subsection{The results for $\Xi_{b}\to \Xi_c$ and $\Xi_{b}\to \Xi^{'}_c$  }

\begin{figure}[hhh]
\begin{center}
\scalebox{0.7}{\includegraphics{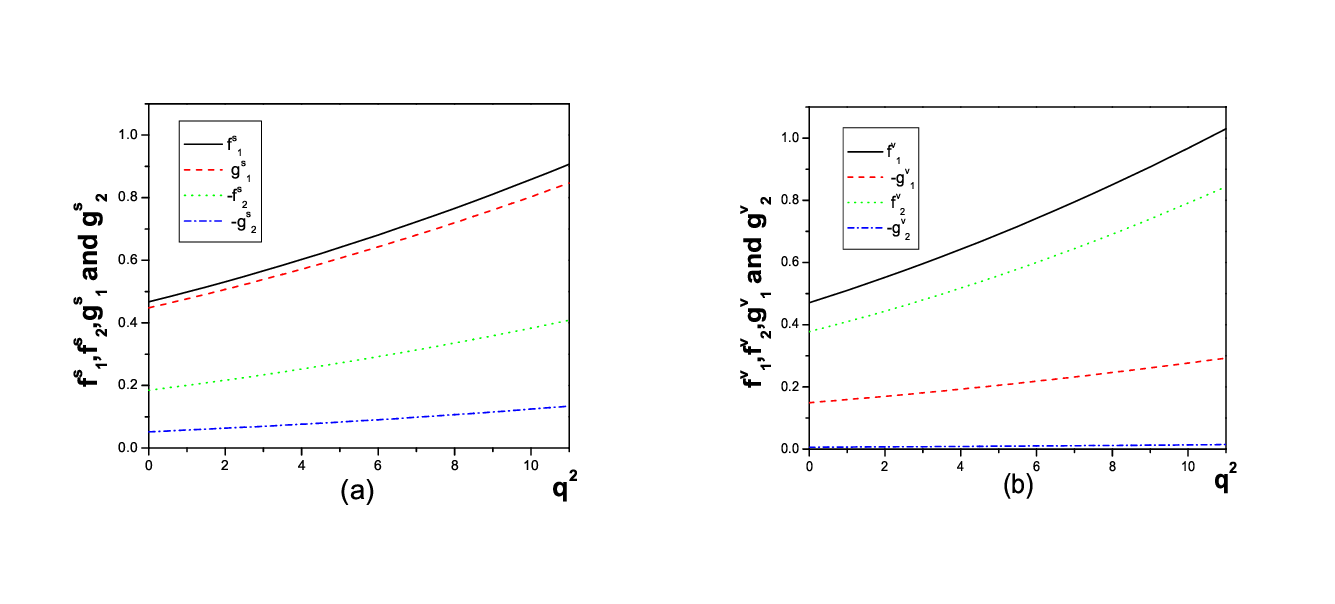}}
\end{center}
\caption{Form factors for the decay
$\Xi^{\bar 3}_{b} \to \Xi^{\bar 3}_{c}$(a) and
$\Xi^{6}_{b} \to \Xi^{6}_{c}$ (b).}\label{fgx}
\end{figure}

After checking the effectiveness of the polynomial we calculate the form factors for $\Xi^{\bar 3}_{b}\to \Xi^{\bar 3}_c$ and $\Xi^6_{b}\to \Xi^{6}_c$ in space-like region and
extrapolate them to time-like region by the Eq. (\ref{p2}). The parameters in the expressions of the form factors are same as those for $\Lambda_b\to\Lambda_c$ in Ref. \cite{Ke:2019smy} except $\beta_{23}\approx 2.9\beta_{us}=0.994$ GeV.
We also depict them in Fig. \ref{fgx}. One can find the form factors for the transition $\Xi^{\bar 3}_{b}\to \Xi^{\bar 3}_c$  are close to those for $\Lambda_b\to\Lambda_c$.

\begin{table}
\caption{The form factors given in
 polynomial form for the transitions $\Xi^{\bar 3}_{b}\to \Xi^{\bar 3}_c$ and $\Xi^6_{b}\to \Xi^{6}_c$ .}\label{xbxcs}
\begin{ruledtabular}
\begin{tabular}{ccccc}
  $F$    &  $F(0)$ &  $a$  &  $b$\\\hline
   $f^s_1$  &   0.467    & 2.19    &2.09   \\
$f^s_2$  &   -0.185     &2.70  &3.01   \\
  $g^s_1$  &      0.448   &    2.09& 1.92  \\
  $g^s_2$  &      -0.052  &    3.32  &  4.54\\
   $f^v_1$  &   0.471    & 2.57    &2.40   \\
$f^v_2$  &   0.378     &2.56  &2.81   \\
  $g^v_1$  &      -0.149    &    2.07& 2.00 \\
  $g^v_2$  &      -0.006  &    2.95  &  2.98
\end{tabular}
\end{ruledtabular}
\end{table}

\begin{table}
\caption{The widths (in unit of $10^{10}{\rm s}^{-1}$) and polarization asymmetries of $\Xi_b\to
\Xi_c l\bar{\nu}_l$ .}\label{xbxcl}
\begin{ruledtabular}
\begin{tabular}{c|ccccc}
    &  $\Gamma$ &  $a_L$  &  $a_T$   & $R$    &  $P_L$ \\\hline
 $\theta=0^\circ$& 3.68 (0.844)&-0.962 (1.03)&-0.745(1.12)&1.63(1.09)&-0.879 (1.06) \\\hline
$\theta=16.27^\circ$
    & 3.13 (0.718)     &   -0.973 (1.04)& -0.756(1.13) & 1.78(1.19) & -0.895(1.08)\\\hline
$\theta=85.54^\circ$
    & 1.15 (0.264)    &   0.685(-0.734) & -0.237(0.355) & 6.69(4.49) & 0.565(-0.684)
  \end{tabular}
\end{ruledtabular}
\end{table}

\begin{figure}[hhh]
\begin{center}
\scalebox{0.7}{\includegraphics{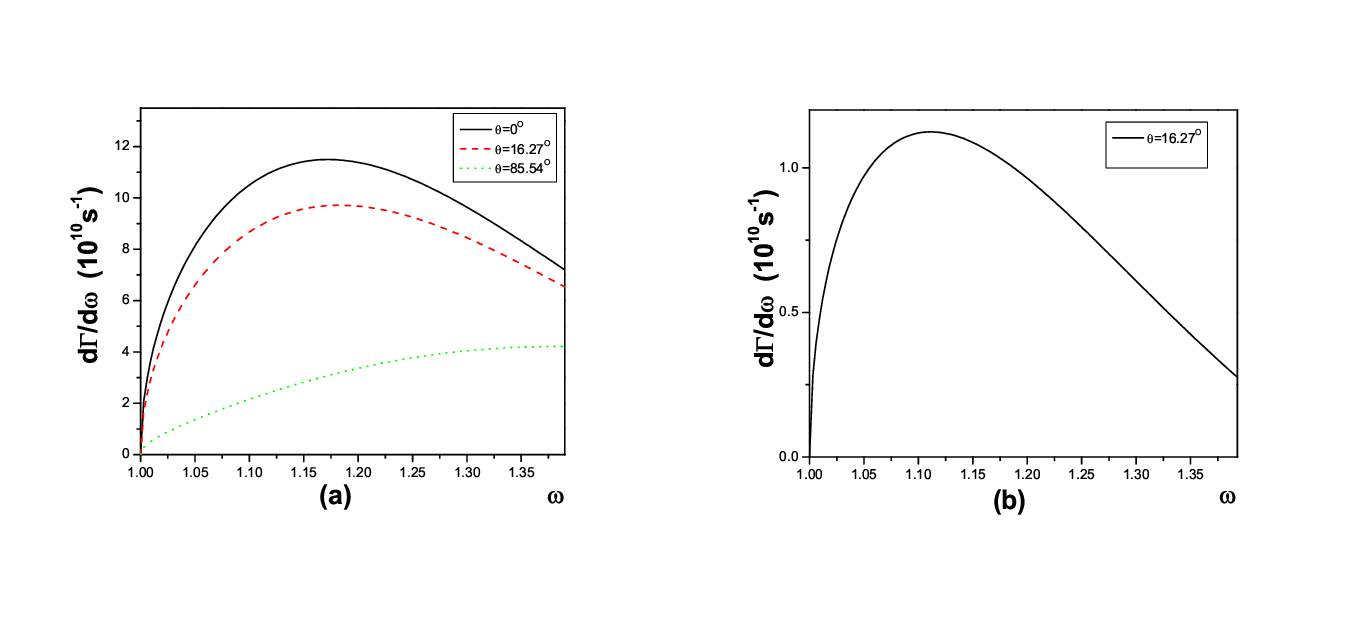}}
\end{center}
\caption{ Differential decay rates $d\Gamma/d\omega$ for the decay
$\Xi_{b} \to \Xi_{c} l\bar{\nu}_l$ (a) and $\Xi_{b} \to \Xi^{'}_{c}
l\bar{\nu}_l$ (b).}\label{dgtheta0}
\end{figure}

Pre-setting different mixing angles, we evaluate the rates and some asymmetries parameters for
the semilepotonic decay and nonleptonic decays of $\Xi_{b} \to \Xi_{c}
$. In tables \ref{xbxcl} and \ref{xbxcn}  we explicitly list the
theoretical predictions with different mixing angles. The dependence of  differential decay widths $d\Gamma/d\omega$
($\omega=\frac{P\cdot P'}{mm'}$) on $\omega$ are depicted in Fig.
\ref{dgtheta0} (a). In order to minimize the model dependence of results,  we also calculate the ratio of these values to those for $\Lambda_{b}
\to \Lambda_{c} l\bar{\nu}_l$ which are presented in the
parentheses in tables \ref{xbxcl}\footnote{In table \ref{xbxcn}, \ref{xbxclp} and \ref{xbxcnp} the value in every
parenthes is the ratio with respect to that of $\Lambda_{b}
\to \Lambda_{c}$}. One may notice when
$\theta=0^\circ$ and $\theta=16.27^\circ$ the polarization
asymmetries of $\Xi_{b} \to \Xi_{c} l\bar{\nu}_l$ are very close to
each other and the decay widths has a slight difference. For a large mixing angle such as $\theta=85.54^\circ$,
the mixing angle apparently changes the decay rate and  polarization
asymmetries relative to the case at $\theta=0^\circ$ in the decay
$\Xi_{b} \to \Xi_{c} l\bar{\nu}_l$. In terms of  $\mathcal{R}(\frac{\Xi^0_b}{\Lambda_b})$,
we find the ratio $\frac{\sigma(\Xi_b^0)}{\sigma(\Lambda^0_b)}\approx 18.4$ when $\theta$ takes $16.27^\circ$.

 For the transition $\Xi_{b}
\to \Xi'_{c} $,  when $\theta=0^\circ$, the theoretical results are
zero in our calculation since the spin of spectator doesn't change. Certainly there maybe exist a very small decay rate for the transition of spin-flip in some other approaches but the situation should be similar to the case of the transition $\Lambda_{b}
\to \Sigma_{c} $ which is very small so that it isn't  listed in PDG databook \cite{ParticleDataGroup:2022pth}.

If the mixing angle $\theta=16.27^\circ$ the semileptonic decay of $\Xi_{b}
\to \Xi'_{c} $ will be one order smaller than that of $\Xi_{b}
\to \Xi_{c}$ and the nonleptonic decay of $\Xi_{b}
\to \Xi'_{c} $ will fifteen to twenty times smaller than that of $\Xi_{b}
\to \Xi_{c}$. The dependence  of
the differential decay widths $d\Gamma/d\omega$
($\omega=\frac{P\cdot P'}{mm'}$) on $\omega$ are depicted in Fig.
\ref{dgtheta0} (b). Comparing  Fig.
\ref{dgtheta0} (a) and (b) one also find the ratio of the differential decay widths at every $\omega$ is also about 10.  In Ref. \cite{LHCb:2023ngz} the signal yields for $\Xi^0_{b}$ and $\Xi^+_{b}$ decay are $462\pm29$ and $175\pm 14$, respectively.  It seems that we have opportunity to observe the decays such as $\Xi_{b}
\to \Xi'_{c}l\nu_l$, $\Xi_{b}
\to \Xi'_{c}D_s$, $\Xi_{b}
\to \Xi'_{c}D^*_s$  and (or) $\Xi_{b}
\to \Xi'_{c}\rho$ in the near future. These processes are very optimal to  determine the mixing angle $\theta$ since $\Xi_{b}
\to \Xi'_{c} $ is very small when $\theta=0^\circ$. Once an observation $\Xi_{b}
\to \Xi'_{c} $ is confirmed, we will have strong evidence to believe that a modest value of mixing angle $\theta$.

In Tables \ref{xbxclp} and \ref{xbxcnp} we only list the results at $\theta=16.27^\circ$. For other mixing angles the decay rates
are different but the polarization asymmetries are same. The results can be understood from the expressions of the form factors $f'_i$
and $g'_i$ in Eq. (\ref{relation1}). To calculate the decay rate of $\Xi_{b}
\to \Xi'_{c} $ for other value of $\theta$, one can  simply re-scale the width in tables \ref{xbxclp} or \ref{xbxcnp}  by a factor ${\rm sin\theta}{\rm cos\theta}/({\rm sin16.27^\circ}{\rm cos16.27^\circ}$).

\begin{table}
\caption{The widths (in unit $10^{10}{\rm s}^{-1}$) and up-down asymmetries of non-leptonic decays
$\Xi_b\to\Xi_c M$.}\label{xbxcn}
\begin{ruledtabular}
\begin{tabular}{c|cc|cc|cc}
  & \multicolumn{2}{c|}{$\theta=0^\circ$~~~~~~}
  & \multicolumn{2}{c|}{ $\theta=16.27^\circ$~~~~} & \multicolumn{2}{c}{ $\theta=85.54^\circ$~~~~}
\\\hline
  & $\Gamma$              & $\alpha$ & $\Gamma$  & $\alpha$   & $\Gamma$  & $\alpha$       \\\hline
  $\Xi_b^0\to\Xi_c \pi^-$  & $0.247$ (0.946)  & $-0.999$ (1.000)
                        & $0.224$ (0.858)  & $-0.988$ (0.989) & $0.143$ (0.548)  & $0.557$ (-0.558)   \\\hline
  $\Xi_b^0\to\Xi_c \rho^-$ & $0.725$ (0.937)  & $-0.877$ (1.002)
                        & $0.651$ (0.841) & $-0.880$ (1.006)& $0.388$ (0.501)  & $0.567$ (-0.648) \\\hline
  $\Xi_b^0\to\Xi_c K^-$    & $0.0198$ (0.943) & $-0.999$ (1.000)
                        & $0.0179$ (0.852) & $-0.987$ (0.988)  & $0.0116$ (0.552) & $0.551$ (-0.552)   \\\hline
  $\Xi_b^0\to\Xi_c K^{*-}$  & $0.0375$ (0.933)  & $-0.839$ (1.002)
                        & $0.0336$ (0.836)  & $-0.846$ (1.011) & $0.0195$ (0.485)  & $0.571$ (-0.682) \\\hline
  $\Xi_b^0\to\Xi_c a_1^-$  & $0.712$ (0.925) & $-0.715$ (1.007)
                     &  $0.641$ (0.832)& $-0.733$ (1.032)  & $0.338$ (0.439) & $0.583$ (-0.821) \\\hline
  $\Xi_b^0\to\Xi_c D_s^-$  & $0.857$ (0.923) & $-0.972$ (0.986)
                        & $0.796$ (0.858) & $-0.941$ (0.954)& $0.612$ (0.659)  & $0.455$ (-0.461) \\\hline
  $\Xi_b^0\to\Xi_c D^{*-}_s$& $1.271$ (0.879) & $-0.338$ (1.024)
                        & $1.090$ (0.754)  & $-0.377$ (1.142) & $0.418$ (0.289)  & $0.643$ (1.948) \\\hline
  $\Xi_b^0\to\Xi_c D^-$  & $0.0330$ (0.924) & $-0.977$ (0.988)
                        & $0.0305$ (0.854)  & $-0.948$ (0.959) & $0.0230$ (0.644)  & $0.466$ (-0.471) \\\hline
  $\Xi_b^0\to\Xi_c {D^*}^-$& $0.0575$ (0.886)& $-0.382$ (1.021)
                        & $0.0495$ (0.763) & $-0.420$ (1.122)& $0.0199$ (0.183)  & $0.636$ (1.701)
\end{tabular}
\end{ruledtabular}
\end{table}

\begin{table}
\caption{The widths (in unit of $10^{10}{\rm s}^{-1}$) and
polarization asymmetries of $\Xi_b\to \Xi'_c l\bar{\nu}_l$
.}\label{xbxclp}
\begin{ruledtabular}
\begin{tabular}{c|ccccc}
    &  $\Gamma$  &  $a_L$  &  $a_T$   & $R$    &  $P_L$  \\\hline
$\theta=16.27^\circ$
    & 0.300(0.0668)      &   -0.0729(0.0781) & -0.622(0.931) & 0.997 (0.669)& -0.348(0.421)
  \end{tabular}
\end{ruledtabular}
\end{table}

\begin{table}
\caption{The widths (in unit $10^{10}{\rm s}^{-1}$) and up-down asymmetries of non-leptonic decays
$\Xi_b\to\Xi'_c M$ with $\theta=16.27^\circ $.}\label{xbxcnp}
\begin{ruledtabular}
\begin{tabular}{ccc}
  & $\Gamma$            & $\alpha$     \\\hline
  $\Xi_b^0\to\Xi'_c \pi^-$  & $0.0142$ (0.0544) & $0.0138$ (-0.0138)
                        \\\hline
  $\Xi_b^0\to\Xi'_c \rho^-$ & $0.0450$ (0.0581) & $0.100$ (-0.114)
                      \\\hline
  $\Xi_b^0\to\Xi'_c K^-$    & $0.00112$ (0.0533) & $0.0180$ (-0.0180)
                          \\\hline
  $\Xi_b^0\to\Xi'_c K^{*-}$  & $0.00238$ (0.0592) & $0.124$  (-0.148)
                         \\\hline
  $\Xi_b^0\to\Xi'_c a_1^-$  & $0.0487$ (0.0632) & $0.194$  (-0.273)
                     \\\hline
  $\Xi_b^0\to\Xi'_c D_s^-$  & $0.0373$ (0.0402) & $0.0902$  (-0.0914)
                        \\\hline
  $\Xi_b^0\to\Xi'_c D^{*-}_s$& $0.107$ (0.0740) & $0.325$  (-0.985)
                       \\\hline
  $\Xi_b^0\to\Xi'_c D^-$  & $0.00149$ (0.0417)& $0.0818$   (-0.0827)
                       \\\hline
  $\Xi_b^0\to\Xi'_c {D^*}^-$& $0.00474$ (0.0730)& $0.317$  (-0.848)
z
\end{tabular}
\end{ruledtabular}
\end{table}

\section{Conclusions and discussions}

In this work we study the transition rates of $\Xi_{b}\to\Xi^{(')}_{c}$ in the light front quark model with a three-quark
picture of baryon. To correctly compute the transition rates,  it is important to properly assess the spin-flavor structures of initial and final states.
In the three-quark picture, the two light quarks constitute a diquark which combine with a heavy quark to form baryons $\Xi_{b}$ and $\Xi_{c}$. In previous studies, the light $qs$ pair in $\Xi_{c}$ ($\Xi^{'}_{c}$ )  was are assumed be to the eigenstates (scalar or pseudovector ) of SU(3) flavor symmetry. However, experimental measurements such as the ratio $\Gamma(\Xi^{++}_{cc}\to\Xi^{+}_{c}\pi)/\Gamma(\Xi^{++}_{cc}\to\Xi^{'+}_{c}\pi)$ \cite{LHCb:2022rpd}  suggest the spin-flavor structures of $\Xi_{c}$ and $\Xi'_{c}$ should be a mixture of $\Xi^{\bar 3}_{c}$  and $\Xi^6_{c}$.
%In fact the symmetry is upset by the difference between the mass of $s$ quark and those of $u$ and $d$ quarks.
Therefore, here we introduce a mixing angle $\theta$ to define the state $|\Xi_{c} \ra$  as  ${\rm cos}\theta\,
| \Xi^{\bar 3}_{c} \ra + {\rm sin}\theta\, | \Xi^{6}_{c} \ra$, and  $| \Xi^{'}_{c} \ra $ as $-{\rm sin}\theta\,| \Xi^{\bar
3}_{c} \ra +{\rm cos}\theta\, |\Xi^{6}_{c} \ra $. Because of heavy quark symmetry,  we can similarly define  $| \Xi_{b} \ra$  as a mixture of states with the same mixing angle: ${\rm cos}\theta \, | \Xi^{\bar	3}_{b} \ra+{\rm sin}\theta \, | \Xi^{6}_{b} \ra$.

Based on proposed spin-flavor structures of $\Xi_{b}$, $\Xi_{c}$ and  $\Xi^{'}_{c}$,  we calculate semileptonic and nonleptonic decay  of $\Xi_{b}\to\Xi_{c}$ with specific value of mixing angle ($\theta$ = $0^\circ$, $16.27^\circ$ and $85.54^\circ$) used in previous analysis \cite{Ke:2022gxm}. We find the decay width and the asymmetry parameter at $\theta=85.54^\circ$ are very different with those without mixing ($\theta=0^\circ$). However for the mixing angle $\theta=16.27^\circ$, only the decay width has a significant decrease relative to the result without mixing.

Following this theoretical framework,  we also evaluate the rates of semileptonic decay and nonleptonic decay of
$\Xi_{b}\to\Xi'_{c}$. Since the form factors $f'_i={\rm cos}\theta {\rm sin}\,\theta(f^v_i- f^s_i)$, $
g'_i={\rm cos}\theta {\rm sin}\,\theta( g^v_i -g^s_i)$, the theoretical decay rates are proportional to the value ${\rm cos}^2\theta {\rm sin}^2\,\theta$, hence we only display results with $\theta=16.27^\circ$ for the purpose of illustration.
Our numerical results indicate that the decay rates of these decays are about ten to twenty times smaller than the values of the transitions $\Xi_{b}\to\Xi_{c}$, which can be feasible to measure experimentally in the near future. Without a mixing of flavor eigenstates,  the rates of the transitions $\Xi_{b}\to\Xi'_{c}$ should be very tiny (about the same order of the transition of $\Lambda_b\to \Sigma_c$). However, a modest mixing angle such as  $\theta=16.27^\circ$ will make the transition rate large enough for measurements, which makes  these channels ideal for studying the mixing of $\Xi_{c}$ and $\Xi'_{c}$ ($\Xi_{b}$ and $\Xi'_{b}$).  We hope measurement of those decay channels of $\Xi_{b}\to\Xi'_{c}$ will be available in the near future, which will help us to test flavor mixing angle  $\theta$ and elucidate the mechanism of decay of heavy baryons.

\section*{Acknowledgement}

This work is supported by the National Natural Science Foundation
of China (NNSFC) under the contract No. 12075167.

\appendix

\end{document}